\documentclass[a4paper,11pt]{article}
\usepackage{pos}

\newcommand{\be}{\begin{equation}}
	\newcommand{\ee}{\end{equation}}
\newcommand{\ba}{\begin{eqnarray}}
	\newcommand{\ea}{\end{eqnarray}}
\newcommand{\baa}{\begin{array}}
	\newcommand{\eaa}{\end{array}}

\newcommand{\tr}{{\text{tr}}}
\newcommand{\SEE}{S_{\text{EE}}}
\newcommand{\vlatt}{N_t N_s^{d-1}}

\let\originalleft\left
\let\originalright\right
\renewcommand{\left}{\mathopen{}\mathclose\bgroup\originalleft}
\renewcommand{\right}{\aftergroup\egroup\originalright}

\newcommand{\e}{\operatorname{e}}

\newcommand{\of}[1]{\left(#1\right)}

\newcommand{\bof}[1]{\biggl(\bigg.#1\bigg.\biggr)}

\newcommand{\sof}[1]{\bigl(\big.#1\big.\bigr)}
\newcommand{\ssof}[1]{(#1)}
\newcommand{\fof}[1]{\left[#1\right]}

\newcommand{\cof}[1]{\left\{#1\right\}}

\newcommand{\bcof}[1]{\biggl\{\bigg.#1\bigg.\biggr\}}

\newcommand{\ssabs}[1]{| #1|}

\renewcommand*\[{\begin{equation}}
\renewcommand*\]{\end{equation}}
\renewcommand*\hat[1]{\widehat{#1}}
\let\oldstackrel\stackrel
\renewcommand*\stackrel[2]{{\scriptstyle\oldstackrel{#1}{#2}}}
\definecolor{emphcol}{RGB}{0,0,0}
\let\oldemph\emph
\renewcommand*\emph[1]{\oldemph{\textcolor{emphcol}{#1}}}
\let\oldstackrel\stackrel
\renewcommand*\stackrel[2]{{\scriptstyle\oldstackrel{#1}{#2}}}

\newcommand*\getscale[1]{%
  \begingroup
    \pgfgettransformentries{\scaleA}{\scaleB}{\scaleC}{\scaleD}{\whatevs}{\whatevs}%
    \pgfmathsetmacro{#1}{sqrt(abs(\scaleA*\scaleD-\scaleB*\scaleC))}%
    \expandafter
  \endgroup
  \expandafter\edef\expandafter#1\expandafter{#1}%
}

\title{Lattice studies of entanglement entropy in $O(N)$ models at finite densities}

\author*{Aatu Rajala}
\author{Niko Jokela}
\author{Tobias Rindlisbacher}

\affiliation{Department of Physics and Helsinki Institute of Physics,\\
P.O.~Box 64, FI-00014 University of Helsinki, Finland}

\emailAdd{aatu.rajala@helsinki.fi}
\emailAdd{niko.jokela@helsinki.fi}
\emailAdd{tobias.rindlisbacher@helsinki.fi}

\abstract{As a characteristic property of all quantum systems, entanglement participates in many important quantum phenomena. In this proceeding, we employ it in the study of quantum field theories at finite density. We incorporate evaluations of entanglement entropy using the replica trick into MC simulations of $O(N)$ models at finite density with the worm algorithm and present some initial results for the nonlinear $O(4)$ model in 3 dimensions.

\textit{Preprint: HIP-2026-3/TH}}

\FullConference{The 42nd International Symposium on Lattice Field Theory (LATTICE2025)\\
2-8 November 2025\\
Tata Institute of Fundamental Research, Mumbai, India\\}

\begin{document}
\maketitle

\section{Introduction}
Entanglement is a fundamental property of quantum systems and responsible for many interesting and important phenomena. The amount of entanglement in a system can be quantified by so-called entanglement measures, for which \emph{entanglement entropy} (EE) is a well known example. For a Hilbert space divided into two subspaces $A$ and $B$, EE can be defined in terms of the reduced density matrix $\rho_A=\tr_B(\rho)$ as 
\begin{equation}
    \SEE(A) = -\tr\left(\rho_A\log \rho_A\right).
\end{equation}
Although this definition is conceptually straightforward, EE is notoriously difficult to compute in interacting quantum field theories (QFT) due to the logarithm. Despite that, there has been significant progress in the lattice computations of EE and other entanglement measures ~\cite{Buividovich:2008kq,Nakagawa:2009jk,Nakagawa:2010kjk,Itou:2015cyu,Rabenstein:2018bri,Rindlisbacher:2022bhe,Jokela:2023rba,Alba:2016bcp,Bulgarelli:2023ofi,Bulgarelli:2024onj,Bulgarelli:2024yrz,Bulgarelli:2025ewp,Amorosso:2024leg,Amorosso:2024glf,amorosso2026} in recent years using the replica trick~\cite{Calabrese:2004eu,Calabrese:2009qy}, which relates the trace of an integer power of the reduced density matrix to a fraction of partition functions.

In this proceeding we harness EE to study the properties of QFTs at finite density. This is done by adapting the boundary deformation method introduced in \cite{Rindlisbacher:2022bhe,Jokela:2023rba} for evaluating a derivative of EE in $SU(N)$ gauge field theories to $O(N)$ models, which can be simulated directly at finite densities with a worm algorithm \cite{prokof2001worm}. 

\section{$O(N)$ models at finite density and the worm algorithm}
The lattice action of a general $O(N)$ model at chemical potential $\mu$ with sources $j$ can be written as 
\begin{align}
S\fof{\phi}\,=\,\sum\limits_{x}\bcof{&-\frac{\kappa}{2}\,\sum\limits_{\nu=1}^{d}\sof{\phi^{\intercal}_{x}\,\e^{2\,\mu\,\tau_{12}\,\delta_{\nu,d}}\,\phi_{x+\hat{\nu}}+\phi^{\intercal}_{x+\hat{\nu}}\,\e^{-2\,\mu\,\tau_{12}\,\delta_{\nu,d}}\,\phi_{x}} \nonumber\\
&+\sof{\phi_{x}\cdot\phi_{x}}+\lambda\sof{\sof{\phi_{x}\cdot\phi_{x}}-1}^{2}-\of{j\cdot\phi_{x}}}\label{eq:onaction},
\end{align}
where $\kappa$ is the hopping parameter, $\lambda$ is a $\phi^4$ coupling, and $\tau_{12}$ is the $N\times N$ matrix
\begin{equation}
    \tau_{12}=\begin{pmatrix} 0 & i & 0 & \cdots \\ -i & 0 \\ 0 & & \ddots  \\ \vdots & & & \ddots\end{pmatrix}.
\end{equation}
We have left the dependence on the lattice spacing $a$ implicit. $\mu$ has also been scaled by a factor of $2$ in comparison to the continuum case.

The action is complex for non-zero $\mu$, which prohibits importance sampling in terms of the $\phi$ fields. This \emph{sign problem} can be circumvented by reformulating the theory in terms of integer-valued dual variables, which can be sampled with a worm algorithm \cite{prokof2001worm}. There are several formulations for the dual variables and the corresponding worm updates. We use the one developed in \cite{Rindlisbacher:2015xku,Rindlisbacher:2016zht,Rindlisbacher:2017ysn}. Alternative approaches and applications beyond the $O(N)$ model can be found in \cite{Gattringer:2012df,Gattringer:2012ap,Endres:2006xu,Bruckmann:2015sua} and \cite{Schmidt:2012uy,DelgadoMercado:2012tte,Bruckmann:2015sua,Rindlisbacher:2016cpj}, respectively.

The partition function written in terms of the dual \emph{flux variables} reads
\begin{align}
Z=\sum\limits_{\cof{k,l,\chi,p,q,n}}\prod\limits_{x}\bcof{&\delta\sof{p_x+\sum_{\nu}\of{k_{x,\nu}-k_{x-\hat{\nu},\nu}}}\times\bof{\prod\limits_{i=3}^{N}\delta_{2}\of{L_{x}^{i}+M_{x}^{i}}}\,\e^{2\mu\,k_{x,d}}\nonumber\\
&\times\bof{\prod\limits_{\nu=1}^{d} w_{\mathrm{l}}\of{L_{x,\nu}; \kappa}}\,w_{\mathrm{s}}\of{L_{x},M_{x}; \lambda, j}}, \label{eq:partition function}
\end{align}
where $k_{x,\nu}\in\mathbb{Z}$ counts the charged net flux from site $x$ to site $x+\hat{\nu}$ on the link $\of{x,\nu}$, $l_{x,\nu}\in\mathbb{N}_0$ is the number of neutral pairs of charged particles, and $\chi_{x,\nu}^{\of{i}}\in\mathbb{N}_0$ the number of neutral particles of type $i$ moving along this link. The \emph{monomers} $p_x\in\mathbb{Z}$, $q_x\in\mathbb{N}_0$, and $n^{\of{i}}_x\in\mathbb{N}_0$ are the total charge, the number of neutral pairs of charged particles, and the number of neutral particles of type $i$ at site $x$, respectively. In \eqref{eq:partition function}, the following abbreviations are also used: $L_{x,\nu}=\ssof{k_{x,\nu},l_{x,\nu},\chi_{x,\nu}^{\of{3}},\ldots,\chi_{x,\nu}^{\of{N}}}$ is a $N$-tuple characterizing the flux variable configuration on the link $\of{x,\nu}$, $L_x$ is a $N$-tuple with components $L^{i}_x=\sum_{\nu=1}^{d}\ssof{\ssabs{L^{i}_{x,\nu}}+\ssabs{L^{i}_{x-\hat{\nu},\nu}}}$, counting the flux of type $i$ on links attached to site $x$, and $M_x=\ssof{p_x,q_x,n_x^{\of{3}},\ldots,n_x^{\of{N}}}$ is a $N$-tuple describing the monomer content on site $x$. The functions $w_l\of{\cdot;\kappa}$ and $w_s\of{\cdot,\cdot;\lambda,j}$ are link and site weights, respectively. 
Finally, there are the on-site constraints imposed by the discrete delta function $\delta\of{\cdot}$, which enforces local charge conservation, and the evenness constraint $\delta_2(\cdot)$, which requires the sum of neutral flux of a given type attached to a site and the corresponding monomer number on that site to be even. These constraints impose severe restrictions on the allowed value combinations of the variables $k$, $\chi^{\of{i}}$, $p$, and $n^{\of{i}}$, rendering sampling with standard local Metropolis updates inefficient. However, the constrained system of flux variables and monomers is ideally suited to be updated with a worm algorithm. The worm algorithm was first introduced in the context of spin models \cite{prokof2001worm} by Prokof'ev and Svistunov as a substitute to cluster algorithms \cite{Swedsencluster} in order to combat critical slowing-down. Here, we will outline the basic idea of how the variables $k$ and $\chi^{\of{i}}$ are sampled, since this is also relevant to our procedure for determining the entanglement measures.

\begin{figure}
    \centering
    \includegraphics[scale=0.5]{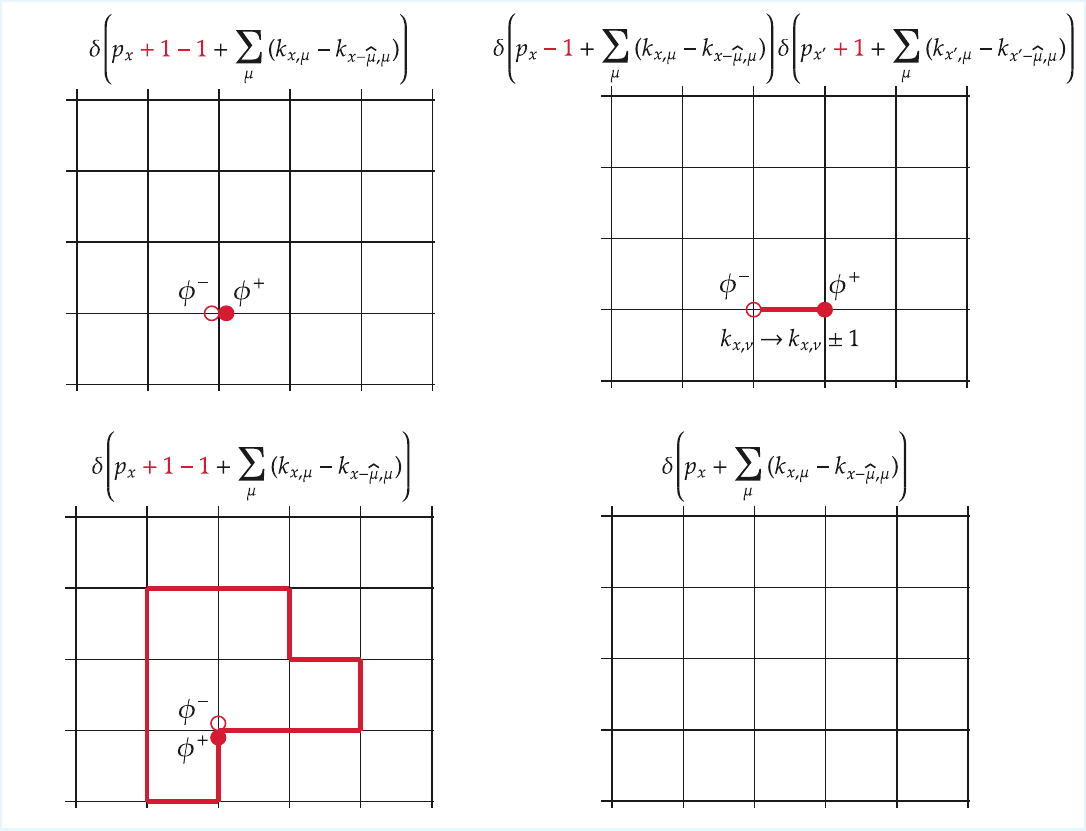} \\[-5pt]
    \caption{An illustration of the worm update in the $k$-sector. The first step is the insertion of the external source-sink pair $\phi^+\phi^-$. Then the head $\phi^+$ is moved to a neighboring site $x'=x+\hat{\nu}$ and $k_{x,\nu}$ is shifted by $\pm1$ (depending on whether $\nu$ is a positive or negative direction) to compensate for the charge displacement. This is continued until the head and tail are again on the same site when they can be removed.} 
    \label{fig:worm_update}
\end{figure}

A single instance of the worm algorithm begins with an insertion of a source and sink pair into the system at some random site $x$. In the $k$ sector this means placing a $\phi^+\phi^-$ pair into the system, which corresponds to adding a $+1$ and a $-1$ into the delta function constraint at $x$
\begin{equation}
    \delta(p_x\textcolor{red}{+\,1-1}+\sum_\mu(k_{x,\mu}-k_{x-\hat{\mu},\mu})).
\end{equation}
As the total change in the expression is zero, the constraint is still satisfied. Next, we propose to move $\phi^+$ to a random neighboring site $x'=x+\hat{\nu}$. In terms of the delta functions, the $+1$ is moved from the delta function at $x$ to the one at $x'$
\begin{equation}
    \delta(p_x\textcolor{red}{-1}+\sum_\mu(k_{x,\mu}-k_{x-\hat{\mu},\mu}))\delta(p_{x'}\textcolor{red}{+1}+\sum_\mu(k_{x',\mu}-k_{x'-\hat{\mu},\mu})).
\end{equation}
Now, the expressions in the delta functions equal $+1$ and $-1$ for $x'$ and $x$ respectively. In other words, the constraints are violated at both sites. However, if simultaneously to the head moving from $x$ to $x'$, we shift the flux variable of the corresponding link by plus or minus one, $k_{x,\nu}\rightarrow k_{x,\nu}\pm1$, depending on whether $\nu$ is a positive or negative direction, the expressions in the delta functions remain zero. In the worm algorithm, the head is moved around the lattice in this way updating the $k$ variables along its path until it and the tail are once again at the same site $x$. Then a removal of the source and sink pair can be attempted, which corresponds to a removal of the $+1$ and $-1$ from the constraint at $x$. Completing a single instance of this entire process is referred to as a single worm update. An illustration of this process is shown in Fig. \ref{fig:worm_update}. The sink $\phi^-$ can also be chosen as the head. Then the $-1$ will move between the constraints and the $k$ variables are updated accordingly.

The update works similarly for the $\chi^{\of{i}}$ sectors. The insertion of the $\phi^i\phi^i$ pair corresponds to adding two $+1$s into the evenness constraints. One of them is then moved around the lattice. Now, as the constraints are simply for the expressions $L^i_x+M^i_x$ in \eqref{eq:partition function} to be even, the $\chi^{\of{i}}$s can be shifted by either $+1$ or $-1$ to keep the constraints satisfied, provided the $\chi^{\of{i}}$ remain non-negative.

\section{Entanglement entropy in lattice $O(N)$ models}
\begin{figure}[htb!]
    \centering
    \includegraphics[width=0.6\textwidth]{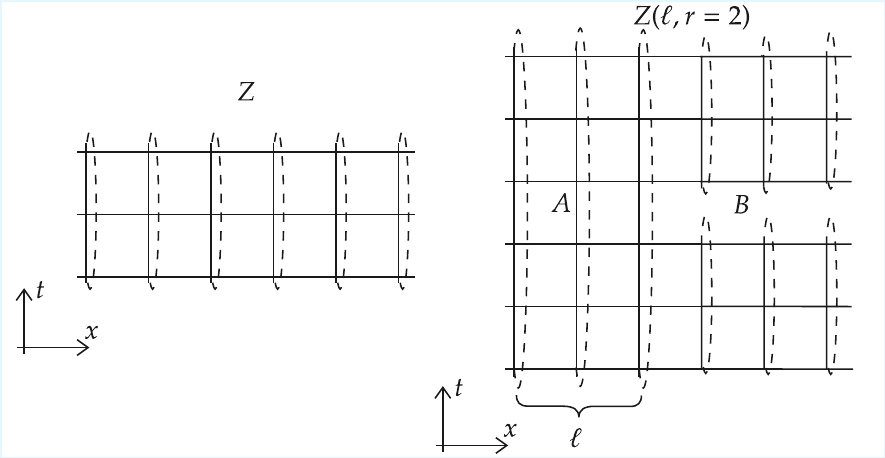}\\[-5pt]
    \caption{Illustration of the topologies of $Z$ and $Z(l,r=2)$. The fields are $1/T$ periodic in the temporal direction for $Z$. For $Z\of{\ell,2}$, in region $A$ of width $\ell$, the fields are $2/T$ periodic. In region $B$, they are $1/T$ periodic. We have chosen to set $r=2$ here to provide a more clear example of $Z\of{\ell,r}$ and because it is the partition function used to approximate $\partial_\ell\SEE$. }
    \label{fig:replica}
\end{figure}

Assuming a slab-shaped entangling region $A$ of width $\ell$, the replica method states that
\begin{equation} \label{eq:replica}
    \text{tr}(\rho^r_A)=\frac{Z(\ell,r)}{Z^r}.
\end{equation}
Here, $Z$ is the standard partition function of field configurations with $1/T$-periodicity in the Euclidean time direction. For $Z(\ell,r)$ the fields are $r/T$ periodic in region $A$, and region $B$ consists of $r$ replicas with $1/T$ periodicity. See Fig. \ref{fig:replica} for visualization.

As EE is UV-divergent, investigating it directly is difficult. However, its derivative with respect to $\ell$ is finite since the divergent part is independent of $\ell$. Calculating this derivative on the lattice usually involves estimating $\SEE$ with the second R\'enyi entropy $H_2\of{\ell}$ ~\cite{Buividovich:2008kq,Rabenstein:2018bri}
\begin{equation}
    H_r\of{\ell}=\frac{1}{1-r}\log\tr\of{\rho^r_A}=\frac{1}{1-r}\of{\log Z\of{\ell,r}-r\log Z}.\label{eq:secrenyientropy}
\end{equation}
Then $\partial_\ell \SEE$ becomes
\begin{equation}
    \left. \frac{\partial \SEE(\ell')}{\partial \ell'} \right\vert_{\ell'=\ell+1/2}\approx -\log Z(\ell+1,2)+\log Z(\ell,2),\label{eq:seeder_latt}
\end{equation}
which is a finite difference approximation of $\partial_\ell H_2$. Increasing $\ell$ by one as required in \eqref{eq:seeder_latt} is a significant non-local change when dealing with more than one spatial dimension. As such, there is minimal overlap between the distribution of configurations contributing to $Z\of{\ell+1,2}$ and the distribution of configurations contributing to $Z\of{\ell,2}$, which presents an issue in importance sampling based lattice simulations.

A way around this issue, introduced in \cite{Rindlisbacher:2022bhe,Jokela:2023rba}, is to deform the boundary between region $A$ and $B$ piece by piece, meaning that the temporal boundary conditions are changed over individual spatial sites at a time. The partition functions $Z\of{\ell,2}$ and $Z\of{\ell+1,2}$ are then connected by an ordered sequence of local boundary deformations. The algorithm can move back and forth along this sequence, while histograms are collected on how often each boundary configuration is visited. $\partial_{\ell}\SEE$ is then evaluated by subtracting the logarithms of these histograms from each other. Statistical uncertainties are estimated with jack-knife resampling.

The constraints in \eqref{eq:partition function} complicate the evaluation of $\partial_\ell \SEE$ with the above procedure. When the boundary conditions over a spatial site are changed, the endpoints of two temporal links are swapped. If the flux variable configuration on these two temporal links are not identical, the swap of endpoints causes the incoming $k$ and $\chi^{\of{i}}$ fluxes on these endpoints to change, leading to \emph{defects}, i.e. violations of the on site constraints on these sites. In the following we describe two alternative procedures we have developed to avoid the formation of defects during boundary deformation updates.

\begin{figure}[htb!]
    \centering
    \includegraphics[width=0.83\textwidth]{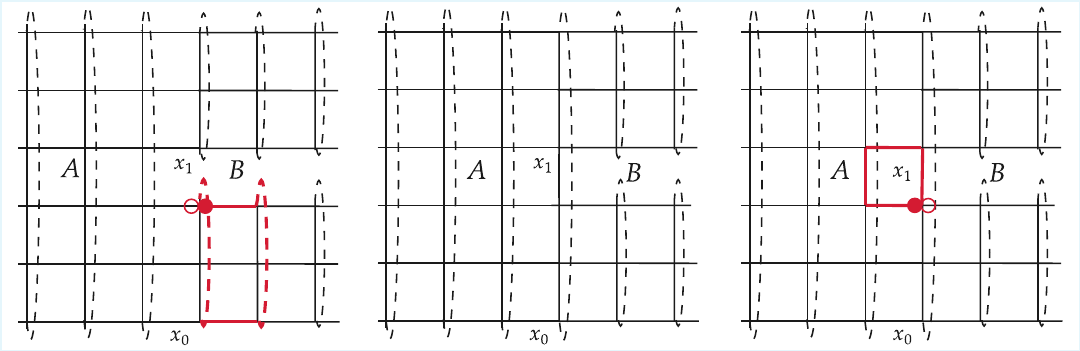}\\[-5pt]
    \caption{A visualization of the plaquette boundary update. Plaquette worms are worm updates that are constrained to move along a temporal plaquette. They change the values of the $k_i=k_{x_i-\hat{t},t}$ and $\chi_i=\chi_{x_i-\hat{t},t}$ such that $\Delta k=k_1-k_0=0$ and $\Delta \chi=\chi_1-\chi_0=0 \ (\text{mod} \ 2)$, which will allow the boundary conditions to be changed without producing defects. After the change, the same plaquettes are done in reverse to restore $\Delta k$ and $\Delta \chi (\text{mod} \ 2)$ to their original values. }
    \label{fig:plaq}
\end{figure}

First, the $k$ and $\chi^{\of{i}}$ configurations on the problematic temporal links can be made compatible with a worm update where the head is constrained to move through a temporal plaquette which contains one of the problematic temporal links (cf. Fig.~\ref{fig:plaq}). The entire boundary update routine performs as many of these \emph{plaquette worms} as necessary to ensure that the boundary change will no longer cause violations of the charge conservation and evenness constraints. Then the boundary conditions can be changed, after which the previously performed plaquette worms are performed in reverse, in order to restore a situation that would trigger the same type of plaquette worms if the reverse boundary update were to be executed next. The latter ensures detailed balance.

\begin{figure}[htb!]
    \centering
    \includegraphics[width=0.83\textwidth]{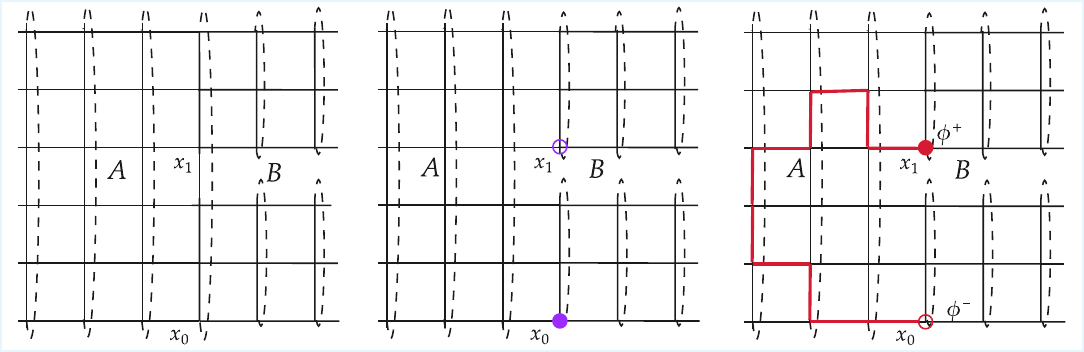}\\[-5pt]
    \caption{A pair of defects introduced by the change of temporal boundary conditions consists of a defect (purple disk) on sites $x_0$ and an anti-defect (purple circle) on site $x_1$ (middle panel). A worm (red) can move the defect from site $x_0$ around till it eventually reaches the site $x_1$ (right panel), after which the defects can be removed.}
    \label{fig:BC_worm}
\end{figure}

The second approach is based on the observation that the defects produced during a change of temporal boundary conditions over a spatial site always appear as defect-anti-defect pairs, with defect and anti-defect being located on different sites. We can therefore use worm updates to move the defect around until it reaches the location of its anti-defect upon which both can be removed (cf. Fig.~\ref{fig:BC_worm}). These worm updates have to be restricted so that they do not change the flux variable configuration on the two temporal links that change endpoints under the change of temporal boundary conditions. This ensures equal selection probability for the reverse boundary update move and enables detailed balance.

\section{Results}
In the following, we report on some results from our simulations of the nonlinear $O(4)$ model in $d=3$ dimensions. The lattices were of size $rN_t\times N_xN_s$ with $r=2$, $N_t=\{5,6,7,8,9,10\}$, $N_x=36$ and $N_s=12$. $\ell$ is along the $x$-direction. We set the hopping parameter to $\kappa=1.2$ and the source $j_3=0.2$. The other sources were set to zero. Simulations were performed for multiple values of $\ell$ and $\mu$ to see how $\partial_\ell \SEE\approx \partial_\ell H_2$ depends on them.

\begin{figure}[htb!]
    \centering
    \includegraphics[width=0.45\textwidth]{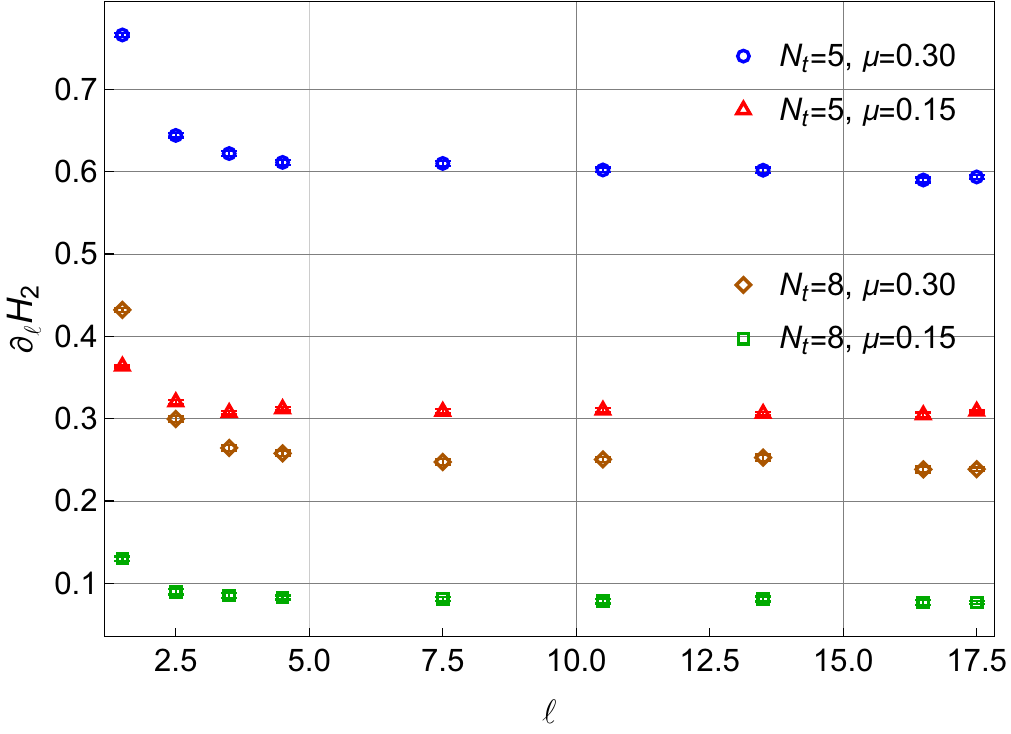}
    \includegraphics[width=0.45\textwidth]{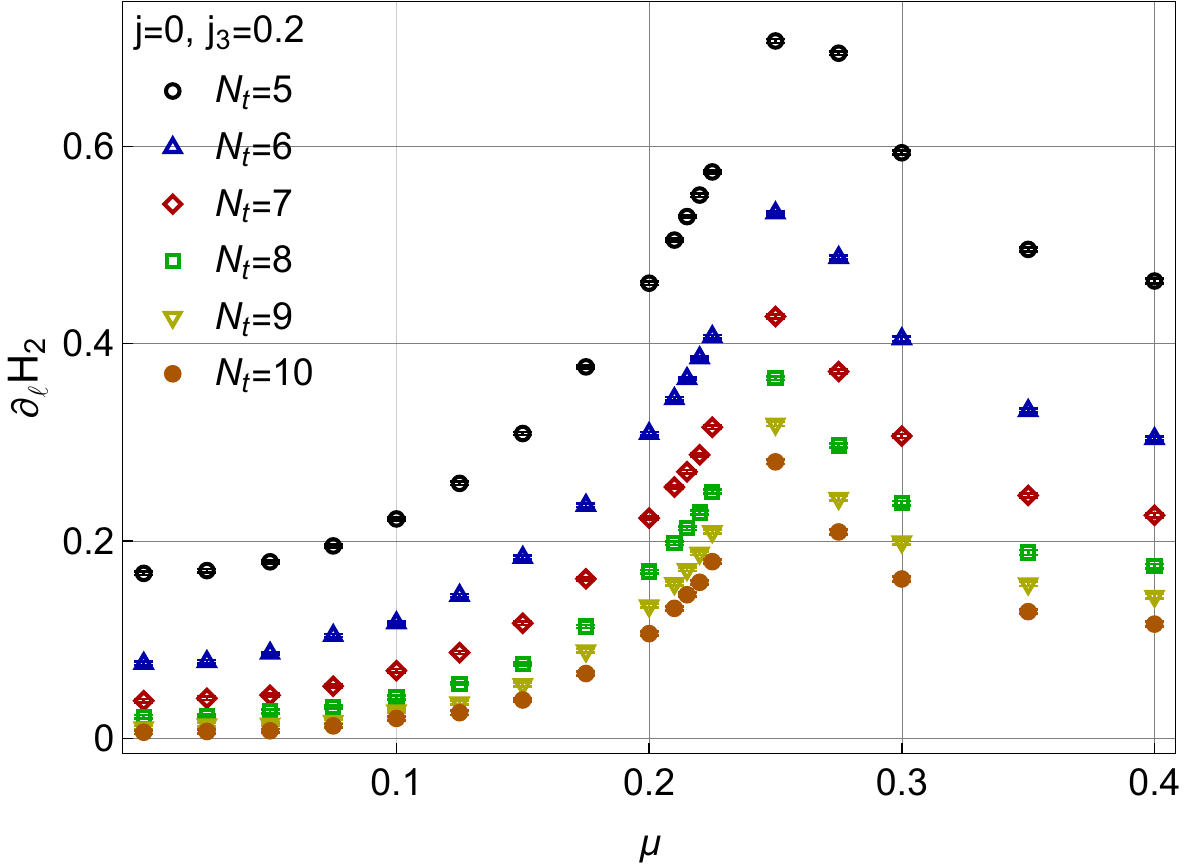}
    \caption{Left: $\partial_\ell H_2\approx\partial_\ell S_{EE}$ as a function of $\ell$ for $N_t=\{5,8\}$ and $\mu=\{0.15;0.30\}$. Right: $\partial_\ell H_2$ as a function of $\mu$ for $N_t=\{5,6,7,8,9,10\}$ and $\ell=17.5$. }
    \label{fig:dSEE_plots}
\end{figure}

The left panel of Fig.~\ref{fig:dSEE_plots} shows $\partial_\ell H_2$ as a function of $\ell$ for $N_t=\{5,8\}$ and $\mu=\{0.15;0.30\}$. All four cases show that $\partial_\ell H_2$ initially decreases rapidly as function of increasing $\ell$ and reaches a $N_t$ and $\mu$  dependent plateau value for $\ell\geq5$.

The $N_t$ and $\mu$ dependence of $\partial_\ell H_2$ is showcased in the right panel of Fig.~\ref{fig:dSEE_plots} where the data is plotted as a function of $\mu$ for $N_t=\{5,6,7,8,9,10\}$ and $\ell=17.5$. It can be seen that as $N_t$ increases $\partial_\ell H_2$ decreases. The relation between $\partial_\ell H_2$ and $\mu$ is more complicated. $\partial_\ell H_2$ increases with $\mu$ until $\mu$ reaches its critical value, after which $\partial_\ell H_2$ decreases.

\begin{figure}[htb!]
    \centering
    \includegraphics[scale=0.38]{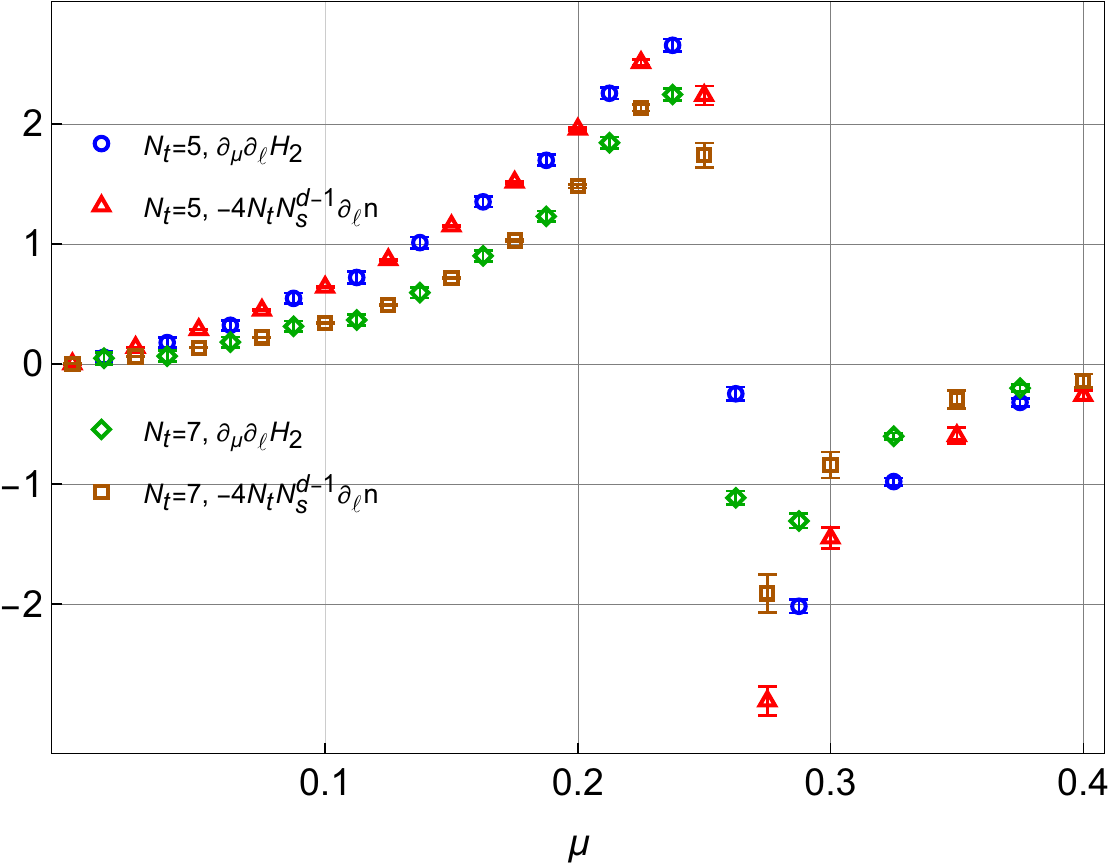}\\[-5pt]
    \caption{Comparison of results for $-\partial_\mu\partial_\ell\log Z\of{\ell,2}$ as computed from $\partial_\mu\partial_\ell H_2$ and $-4\vlatt \partial_\ell n$ for $N_t=\{5,7\}$ and $\ell=17.5$. Both cases show a clear agreement between the two evaluations.}
    \label{fig:derivatives}
\end{figure}

From \eqref{eq:secrenyientropy}, it is apparent that 
\begin{equation}
\partial_\ell H_2 = -\partial_\ell \log Z\of{\ell,2}\ .
\end{equation} 
Therefore, if $\partial_\ell H_2$ is evaluated at different $\mu$ and a numerical derivative with respect to $\mu$ is taken, estimates of the quantity $-\partial_\mu\partial_\ell\log Z\of{\ell,2}$ are obtained. 
Another important quantity obtained via a derivative of this partition function is the charge density,
\begin{equation} \label{eq:lattice_n}
    n\of{\ell,2}=\frac{1}{4\vlatt}\frac{\partial \log Z\of{\ell,2}}{\partial\mu}=\frac{\langle \sum_xk_{x,d}\rangle}{2\vlatt}.
\end{equation}
By measuring the expectation value of $n$ for each boundary condition state used to interpolate between $Z\of{\ell,2}$ and $Z\of{\ell+1,2}$ in the computation of $\partial_\ell H_2$, we can numerically estimate the $\ell$-derivative of $n\of{\ell,2}$, i.e. $\frac{1}{4\vlatt}\frac{\partial^2\log Z\of{\ell,2}}{\partial\ell\partial\mu}$. With a change in the order of derivation and some multiplications, it becomes apparent that
\begin{equation} \label{eq:d2F}
    \frac{\partial^2 H_2}{\partial\mu\,\partial\ell}=-4\vlatt \frac{\partial n\of{\ell,2}}{\partial\ell}. 
\end{equation}
Using the aforementioned procedures, we can check this connection with our lattice simulations testing our algorithm in the process. The results of these evaluations are shown in Fig. \ref{fig:derivatives} which shows plots of $\partial_\mu\partial_\ell H_2$ and $-4\vlatt\partial_\ell n$ as a function of $\mu$ for $N_t=\{5,7\}$ and $\ell=17.5$. The plots match quite well and \eqref{eq:d2F} seems to hold. We have tested the connection in many simulations with a multitude of different parameters in addition to the ones presented in this proceeding and it has held in all of them. This strongly indicates that our algorithm is working as intended.

\section{Conclusion and outlook}
In this proceeding, we have introduced a modified boundary deformation method for measuring $\partial_\ell \SEE$ in lattice $O(N)$ models at finite $\mu$ and presented results for the nonlinear $O(4)$ model in $d=3$. Our results highlight the usefulness of entanglement entropy in the study of phase transitions. Along this line, we plan to compute critical exponents \cite{Pelissetto:2000ek} through entanglement entropy. Comparing simulation results at high $N$ to $N\to\infty$ calculations \cite{Metlitski:2009iyg} would also be interesting. Our algorithm could also be used to simulate canonical ensembles, enabling the study of symmetry-resolved entanglement \cite{Goldstein:2017bua}. In \cite{Jokela:2023rba}, it was argued that in certain limits the large $\ell$ value of $\partial_\ell \SEE$ matches the thermal entropy density. We aim to further establish and test this connection in future publications.

\acknowledgments
A.~R. acknowledges the financial support of the Finnish Ministry of Education and Culture through the Quantum Doctoral Education Pilot Program (QDOC VN/3137/2024-OKM-4) and the Research Council of Finland through the Finnish Quantum Flagship project (358878, UH). N.~J. was supported in part by the Research Council of Finland through grant no. 3545331 and the Centre of Excellence in Neutron-Star Physics (project 374062). T.~R. acknowledges support form the European Research Council grant 101142449 and Research Council of Finland grant 354572. The authors thank CSC - IT Center for Science, Finland, for computational resources. 
\bibliographystyle{JHEP}
\bibliography{refs.bib}

\end{document}